\documentclass[12pt]{article}
\usepackage[dvips]{graphicx}
\usepackage{epsfig}

\def\gev{\,{\rm GeV}}

\newcommand{\nn}{\nonumber}
\newcommand{\be}{\begin{equation}}
\newcommand{\ee}{\end{equation}}
\newcommand{\ba}{\begin{eqnarray}}
\newcommand{\ea}{\end{eqnarray}}
\newcommand{\lsim}{\raisebox{-4pt}{$\,\stackrel{\textstyle
                                                         <}{\sim}\,$}}
\newcommand{\gsim}{\raisebox{-4pt}{$\,\stackrel{\textstyle
                                                         >}{\sim}\,$}}

\def\qbq{{q}\overline{q}}
\def\sbs{{s}\overline{s}}
\def\cbc{{c}\overline{c}}
\def\ubu{{u}\overline{u}}
\def\dbd{{d}\overline{d}}
\def\jp{J/\psi}
\def\mjp{M_{J/\psi}}

\def\als{\alpha_s}
\def\sudrei{{\rm SU}(3)_{{\rm F}}}
\def\uaeins{{\rm U}_{\rm A}(1)}
\def\qb{\overline{Q}}
\def\qqb{\overline{Q}\hspace{1pt}^2}
\begin{document}
\thispagestyle{empty}
\begin{flushright}
WU B 02-01 \\
PITHA 02-01 \\
hep-ph/0201044\\
January 2002\\[5em]
\end{flushright}

\begin{center}
\end{center}
\begin{center}{\Large\bf Mixing of Pseudoscalar Mesons} \\
\vskip 3\baselineskip

Th.\ Feldmann$^a$\,\footnote{Email: feldmann@physik.rwth-aachen.de }
and P.\ Kroll$^b$\,\footnote{Email: kroll@theorie.physik.uni-wuppertal.de}\\[0.5em]
a) {\small {\it Institut f\"ur Theoretische Physik E, RWTH Aachen,
    52056 Aachen, Germany}}\\
b) {\small {\it Fachbereich Physik, Universit\"at Wuppertal, 42097
Wuppertal, Germany}}\\
\vskip \baselineskip
\end{center}
\begin{abstract}
$\eta$ -- $\eta'$ mixing is discussed in the quark-flavor basis with
the hypothesis that the decay constants follow the pattern
of particle state mixing. On exploiting the divergences of the axial
vector currents -- which embody the axial vector anomaly -- all 
mixing parameters are fixed to first order of flavor symmetry breaking. 
An alternative set of parameters is obtained from a phenomenological 
analysis. We also discuss mixing in the octet-singlet basis
and show how the relevant mixing parameters are related to those
in the quark-flavor basis. The dependence of the mixing parameters on
the strength of the anomaly and the amount of flavor symmetry breaking is
investigated. Finally, we present a few applications of the 
quark-flavor mixing scheme, such as radiative decays of vector mesons, 
the photon-pseudoscalar meson transition form factors, the coupling 
constants of $\eta$ and $\eta'$ to nucleons, and the 
isospin-singlet admixtures to the $\pi^0$ meson.
\end{abstract}
 
\begin{center}
Invited talk given by P. Kroll at the Workshop on Eta Physics\\
Uppsala, October, 25-27, 2001
\end{center}

\begin{center}
\end{center}

\begin{center}
\vskip \baselineskip
\end{center}

\newpage
\section{Introduction}
$\eta$ -- $\eta'$ mixing is a subject of considerable interest that has
been examined in many phenomenological investigations, see, e.g.,
\cite{Isgur}--\cite{Gilman}. New aspects of mixing,
which mainly concern the proper definition of meson decay constants 
and the consistent extraction of mixing parameters from experimental
data, have recently been discussed by Kaiser and Leutwyler \cite{kai98} 
and by us \cite{FKS1,FKS2,fel00}. The purpose of the present article
is to review these new developments and to clarify the interplay
between the $\uaeins$ anomaly and flavor symmetry breaking. 
We will not comment on the question of how the $\uaeins$ anomaly 
actually arises in QCD~\cite{tHooft}--\cite{Weinberg:1975ui}.

We start from the quantum mechanical picture of mixing as a
superposition of basis states. This mixing is either described in the
octet-singlet basis, e.g.\ 
\cite{Dono}--\cite{kai98},
\begin{equation}
 \left (\matrix{\eta \cr \eta'}\right )\,=\, U(\theta)\;
                      \left (\matrix{\eta_{8} \cr \eta_{0}}\right ) ,
\label{osb}
\end{equation}
or in the quark-flavor basis, e.g.\ \cite{Isgur}--\cite{Eides},\cite{FKS1},
\begin{equation}
\left (\matrix{\eta \cr \eta'}\right )\,=\, U(\phi)\;
                      \left (\matrix{\eta_{q} \cr \eta_{s}}\right ) ,
\label{qsb}
\end{equation}
where $U$ is a unitary matrix defined as  
\begin{equation}
    U(\alpha)\,=\,\left(\matrix{\cos{\alpha} & -\sin{\alpha} \cr
                                \sin{\alpha} & 
\phantom{-}\cos{\alpha}} \right) .
\label{uni}
\end{equation}
The basic states, $\eta_8$, $\eta_0$, or $\eta_q$, $\eta_s$, are
assumed to be orthogonal states, i.e.\ mixing with heavier pseudoscalar
mesons (e.g.\ possibly glueballs) is ignored. 
They are furthermore assumed to be identifiable
by their valence quark content which are either the
$\sudrei$ octet and singlet combinations of quark-antiquark pairs or,
for the $\eta_q$ and $\eta_s$, the combination $\qbq = (\ubu +
\dbd)/\sqrt{2}$ and $\sbs$, respectively.  We stress
that as long as state mixing is regarded, one may freely transform
from one orthogonal basis to the other. The respective mixing angles
are related to each other by $\theta = \phi - \theta_{\rm ideal}$ where
$\theta_{\rm ideal}= \arctan{\sqrt{2}}$ is the ideal mixing angle. 

The phenomenological analyses of
decay or scattering processes often involve weak decay constants 
of $\eta$ and $\eta'$ mesons which are defined by
\begin{equation}
\langle 0 | J_{\mu5}^i | P (p)\rangle = i \,
f_P^i \, p_\mu \,, \qquad (i=8,0,q,s; \quad P=\eta, \eta')\,.
\label{dec}
\end{equation}
Occasionally, it is assumed that the octet and singlet decay constants,
follow the pattern of state mixing
\begin{eqnarray}
&& f_{\eta\phantom{'}}^8 = f_8 \, \cos\theta \ , \qquad
   f_{\eta\phantom{'}}^0 = - f_0 \, \sin\theta \ , \nonumber \\[0.2em]
&&  f_{\eta'}^8 = f_8 \, \sin\theta \ , \qquad
   f_{\eta'}^0 = \phantom{-} f_0 \, \cos\theta \,.
\label{oldmix}
\end{eqnarray}
However, a recent study within chiral perturbation theory
\cite{kai98} as well as a combined phenomenological analysis of the
meson-photon transition form factors, the two-photon decay widths and
radiative $\jp$ decays \cite{fel97} revealed that (\ref{oldmix}) is
inadequate and theoretically inconsistent. 
The general parameterization \cite{kai98}
\begin{eqnarray}
&& f_{\eta\phantom{'}}^8 =  f_8 \, \cos\theta_8 \ , \qquad
    f_{\eta\phantom{'}}^0 = - f_0 \, \sin\theta_0 \ ,\nonumber \\[0.2em]
&&  f_{\eta'}^8 = f_8 \, \sin\theta_8 \ , \qquad
   f_{\eta'}^0 = \phantom{-} f_0 \, \cos\theta_0 \ ,
\label{newmix}
\end{eqnarray}
is required. According to \cite{kai98,fel97} the angles
$\theta_8$ and $\theta_0$ differ considerably as a consequence of
flavor symmetry breaking.

Analogously to (\ref{oldmix}), one may assume that the
strange and non-strange decay constants follow the
pattern of state mixing in the quark-flavour basis \cite{FKS1}
\begin{eqnarray}
&& f_\eta^{q} = f_{q} \, \cos\phi \ , \qquad
    f_\eta^{s} = - f_{s} \, \sin\phi \ ,\nopagebreak \nonumber \\[0.2em] 
&&  f_{\eta'}^{q} = f_{q} \, \sin\phi \ , \qquad
   f_{\eta'}^{s} =  \ \ f_{s} \, \cos\phi \ .
\label{phimix}
\end{eqnarray}
Alternatively, one may introduce the analogue of (\ref{newmix}) with
two mixing angles, $\phi_q$ and $\phi_s$, here as well. The
phenomenological analysis carried through in \cite{fel97} provided 
$\phi_q\simeq 39.4$, and $\phi_s\simeq 38.5$. The closeness of the two 
angles is sufficiently suggestive to ignore the little difference and
to assume $\phi_q=\phi_s=\phi$ \cite{FKS1}. A theoretical
explanation for this fact is given by the OZI-rule which
implies that the difference between $\phi_q$ and $\phi_s$ vanishes
to leading order in the $1/N_c$ expansion \cite{kai98,fel00}.
For consistency, we neglect all OZI-rule violating effects in
the following.
The decay constants $f_P^q$ and $f_P^s$ then respect the simple 
mixing behaviour (\ref{phimix}) which is equivalent to the hypothesis 
that both the basic states, $\eta_q$ and $\eta_s$, have vanishing 
vacuum transition matrix elements with opposite currents
\be
\langle 0 \mid J^s_{\mu 5} \mid \eta_q \rangle = 
                \langle 0 \mid J^q_{\mu 5} \mid \eta_s \rangle = 0\,.
\label{opposite}
\ee
The analogous relation in the octet-singlet basis does not hold.
With regard to this central assumption the quark-flavor basis is
distinct. It implies among other thinks that the decay constants of
the mesons can be written as simple mass-independent superpositions of
$f_q$ and $f_s$. We will explain in the following sections that the proper
use of this basis provides new insights in $\eta$--$\eta'$ mixing and 
successful predictions.

In Sect.~\ref{sec:qsmix} we will present some technical details 
of the quark-flavor basis and discuss the determination of the basic 
mixing parameters $f_q$, $f_s$ and $\phi$, their relations to the mixing 
parameters, $f_8$, $f_0$, $\theta$, $\theta_8$ and $\theta_0$, 
in the octet-singlet basis, and the 
important role of the  matrix elements of the anomaly operator. 
Next, in Sect.~\ref{sec:a2}, we study the dependence of 
masses and mixing angles on the input parameters that classify the 
strength of the $\uaeins$ anomaly and the flavor symmetry breaking.
Some phenomenological applications of the quark-flavor mixing
scheme will be discussed in Sect.~\ref{sec:appl}. 
Our article ends with a summary (Sect.~\ref{sec:summ}).  

\section{The quark-flavor mixing scheme}
\label{sec:qsmix}

\begin{figure}[tb]
\begin{center}
\psfig{file=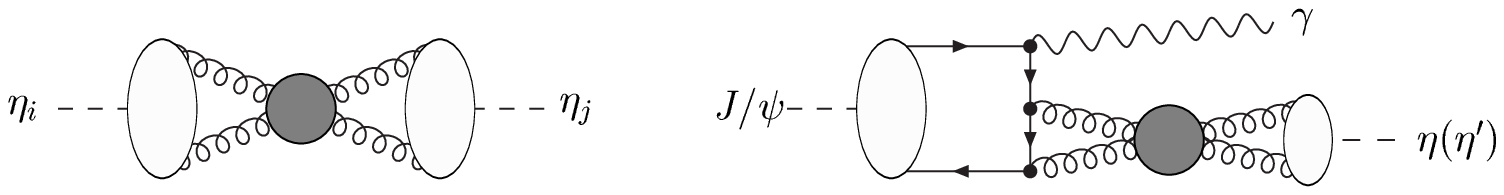, bb= 70 695 245 755, width=0.40\textwidth,clip=}
\end{center}
\caption{$\eta_i$--$\eta_j$ transitions through the $\uaeins$ anomaly
   (indicated by the grey blob).}
\label{fig:anomaly}
\end{figure}

As is well-known \cite{Witten}, the $\uaeins$ axial vector anomaly,
that plays a crucial role for understanding the mixing behaviour of the
pseudoscalar mesons, is embodied in the divergences of axial
vector currents,  
\be
\partial^\mu J_{\mu 5}^{u} =
\partial^\mu (\bar{u} \, \gamma_\mu \gamma_5 \, u) =
2 \, m_{u} \, (\bar{u} \, i \gamma_5 \, {u}) + \frac{\alpha_s}{4\pi} \,
G \, \widetilde G \, .
\ee
Analogous relations hold for the other quark flavors.
Here, $G$ denotes the gluon field strength tensor and $\widetilde G$ its
dual; $m_i$ is the current mass of quark species $i$. The
vacuum-meson transition matrix elements of the axial vector current 
divergences are given by the product of the square of the meson mass,
$M_P^2$, and the appropriate decay constant
\begin{equation}
\langle 0 | \partial^\mu J_{\mu 5}^{i} | P \rangle = M_P^2 \,
                                                          f_P^{i} \,.
\label{div}
\end{equation}
The mass factors, which necessarily appear quadratically here,
can be viewed as the elements of the physical particle mass matrix
\begin{equation}
{\mathcal M}^2 = \left(\matrix{ M_\eta^2 & 0 \cr 0 & M_{\eta'}^2 } \right)
                                                                         \,.
\end{equation} 
Transforming to the quark-flavor basis and exploiting the relations
(\ref{div}), one finds
\begin{equation}
{\mathcal M}^2_{qs} = U^\dagger(\phi) \, {\mathcal M}^2 \, U(\phi) =
\left(\matrix{
             m_{qq}^2  + 2 a^2 & \sqrt{2} y a^2 & \vspace{0.3em}\cr
                 \sqrt{2} y a^2& m_{ss}^2 + y^2 a^2 } \right)\,.
\label{qsmass}
\end{equation} 
The quark mass contributions to ${\mathcal M}^2_{qs}$ are defined as
\be
 m_{qq}^2 = \frac{\sqrt2}{f_{q}} \, 
      \langle 0 | m_{u} \, \bar{u} \, i \gamma_5 \, {u} +
                  m_{d} \, \bar{d} \, i \gamma_5 \, {d} | \eta_{q}
                  \rangle\,,
\qquad
m_{ss}^2 = \frac{2}{f_{s}} \, 
      \langle 0 | m_{s} \, \bar{s} \, i \gamma_5 \, {s} | \eta_{s}
                  \rangle \, ,
\label{miidef}
\ee
and $a^2$ parameterizes the anomaly contribution to the mass matrix,
\begin{equation}
a^2 = \frac{1}{\sqrt2 \, f_{q}} \, \langle 0 | 
                  \frac{\alpha_s}{4\pi} \, G \widetilde G | \eta_{q}
                  \rangle\,.
\label{aadef}
\end{equation}
The anomaly mediates $\eta_q\leftrightarrow \eta_s$ transitions
(see Fig.\ \ref{fig:anomaly}) and therefore leads to $\eta$--$\eta'$
mixing. The vacuum--$\eta_i$ matrix elements of the anomaly operator 
$G\widetilde{G}$, are non-zero and in fact large because of
the non-trivial properties of the QCD vacuum -- there are strong
gluonic fluctuations with pseudoscalar quantum numbers  
to which the $\eta_i$ states can couple.

The parameter $y=f_q/f_s$ measures the strength of flavor
symmetry violation encoded in the decay constants. The symmetry of the mass
matrix forces an important connection between $y$ and anomaly matrix elements
\begin{equation}
y = \frac{f_{q}}{f_{s}}\,=  \sqrt2 \, \frac{ \langle 0 | 
                  \frac{\alpha_s}{4\pi} \, G \widetilde G | \eta_{s} \rangle}
       {\langle 0 | 
                  \frac{\alpha_s}{4\pi} \, G \widetilde G | \eta_{q} \rangle}\,
\ .
\label{ydef}
\end{equation}

\subsection{Determination of the basic mixing parameters}
Eq.\ (\ref{qsmass}) provides three relations which allow the
determination of $a^2$, $y$ and $\phi$ for given masses of the
physical mesons and quark mass terms:
\ba
 \sin {\phi} &=&\; \sqrt{\frac{(M^2_{\eta'}-m^2_{ss})\,(M^2_{\eta}-m^2_{qq})}
               {(M^2_{\eta'}-M^2_{\eta})\,(m^2_{ss}-m^2_{qq})}}\;, \nn\\[0.3em]
 a^2 &=& \;\frac12\; \frac{(M^2_{\eta}-m^2_{qq})\,(M^2_{\eta'}-m^2_{qq})}
                        {(m^2_{ss}-m^2_{qq})} \;, \nn\\[0.3em]
 y&=& \sqrt{2\,\frac{(M^2_{\eta'}-m^2_{ss})\,(m^2_{ss}-M^2_{\eta})}
                            {(M^2_{\eta'}-m^2_{qq})\,(M^2_{\eta}-m^2_{qq})}}\,.
\label{mix-par}
\ea
In order to determine the mixing parameters from (\ref{mix-par}) we
take recourse to first order of flavor symmetry breaking and relate 
the quark mass terms to the pion and kaon masses which themselves 
are not affected by the anomaly
\begin{equation}
m_{qq}^2 = M_\pi^2 \,, \qquad
m_{ss}^2 = 2  M_K^2 - M_\pi^2 \,.
\label{masses}
\end{equation}
Inserting these values into (\ref{mix-par}) one gets - to the given
order - parameter-free results for the mixing parameters which are
quoted in Tab.\ \ref{tab:par}. To the same order of flavor symmetry
breaking one also has the theoretical estimate 
\begin{equation}
f_{q} = f_\pi \,, \qquad
f_{s} =  \sqrt{2  f_K^2 - f_\pi^2}\,.
\label{phen6}
\end{equation}
\begin{table}[t]
\begin{center}
\begin{tabular}{|c| ccccc| }
\hline
source & $f_{q}/f_\pi$ & $f_{s}/f_\pi$ & $\phi$ & $y$
& $a^2$ $[$GeV$^2]$ \\
\hline\hline
theory  & 
 $\protect\phantom{\pm}1.00$ & $\protect\phantom{\pm}1.41$ & 
 $\protect\phantom{\pm}42.4^\circ$ &  
$\protect\phantom{\pm}0.78$ & $\protect\phantom{\pm}0.281$ \\
\hline
phenomenology   &
 $\protect\phantom{\pm}1.07$ & $\protect\phantom{\pm}1.34$ & 
 $\protect\phantom{\pm}39.3^\circ$ &  
 $\protect\phantom{\pm}0.81$ & $\protect\phantom{\pm}0.265$ \\
 & $\pm 0.02$ & $\pm 0.06$ & $\pm \protect\phantom{3}1.0^\circ$ & 
 $\pm 0.03$ & $\pm0.010 $ 
\\ \hline
\end{tabular}
\end{center}
\caption{Theoretical (to first order of flavor symmetry
breaking) and phenomenological values of
mixing parameters. The parameter $y$ is calculated using
(\ref{mix-par}). The error estimate refers to the experimental
uncertainties only. Table taken from \cite{FKS1}.}
\label{tab:par}
\end{table}

One may also determine the mixing parameters from phenomenology and
look for consistency with or deviations from the first order
theoretical results. The mixing angle $\phi$ can cleanly be extracted
from a number of processes which have been analyzed in Refs.\
\cite{FKS1,BaFrTy95,Bramon97}. The idea is to consider {\em ratios}
of $\eta$ and $\eta'$ observables in which the dependence on
form factors, decay constants etc.\ cancels, and only functions of the
mixing angle (times kinematical pre-factors) appear. 
In Tab.\ \ref{tab:phi} we compile the results of a detailed 
phenomenological analysis presented in \cite{FKS1}. It yields a
weighted average for the mixing angle $\phi$  
\begin{eqnarray} 
\phi_{av} = 39.3^\circ \pm 1.0^\circ.
\label{ave}
\end{eqnarray}
where the error refers to the experimental uncertainties, only. 
Quite remarkably, the values for $\phi$ obtained from very different 
physical processes are all compatible with each other within the
errors. This would not be the case in the octet-singlet scheme (if  
$\theta_8=\theta_0=\theta$ were assumed). This fact used to be a 
known problem in previous analyses (for instance 
\cite{Dono,Gilman,BaFrTy95,GaLe85}, see also Tab.~\ref{tab:comp})
where values for $\phi$ varying  from $\simeq 45^\circ$ to 
$\simeq 32^\circ$ (corresponding to $-10^\circ\, \lsim
\theta \lsim \,-23^\circ$) have been found. The phenomenological 
value of $\phi$ in (\ref{ave}) does not differ substantially from 
the theoretical value quoted in Tab.\ \ref{tab:par}, i.e.\ systematic 
effects from higher order flavor symmetry breaking corrections etc., 
are apparently not large. 

\begin{table}[t]
\begin{center}
\fbox{\parbox{10.2cm}{\epsfclipon
\psfig{file=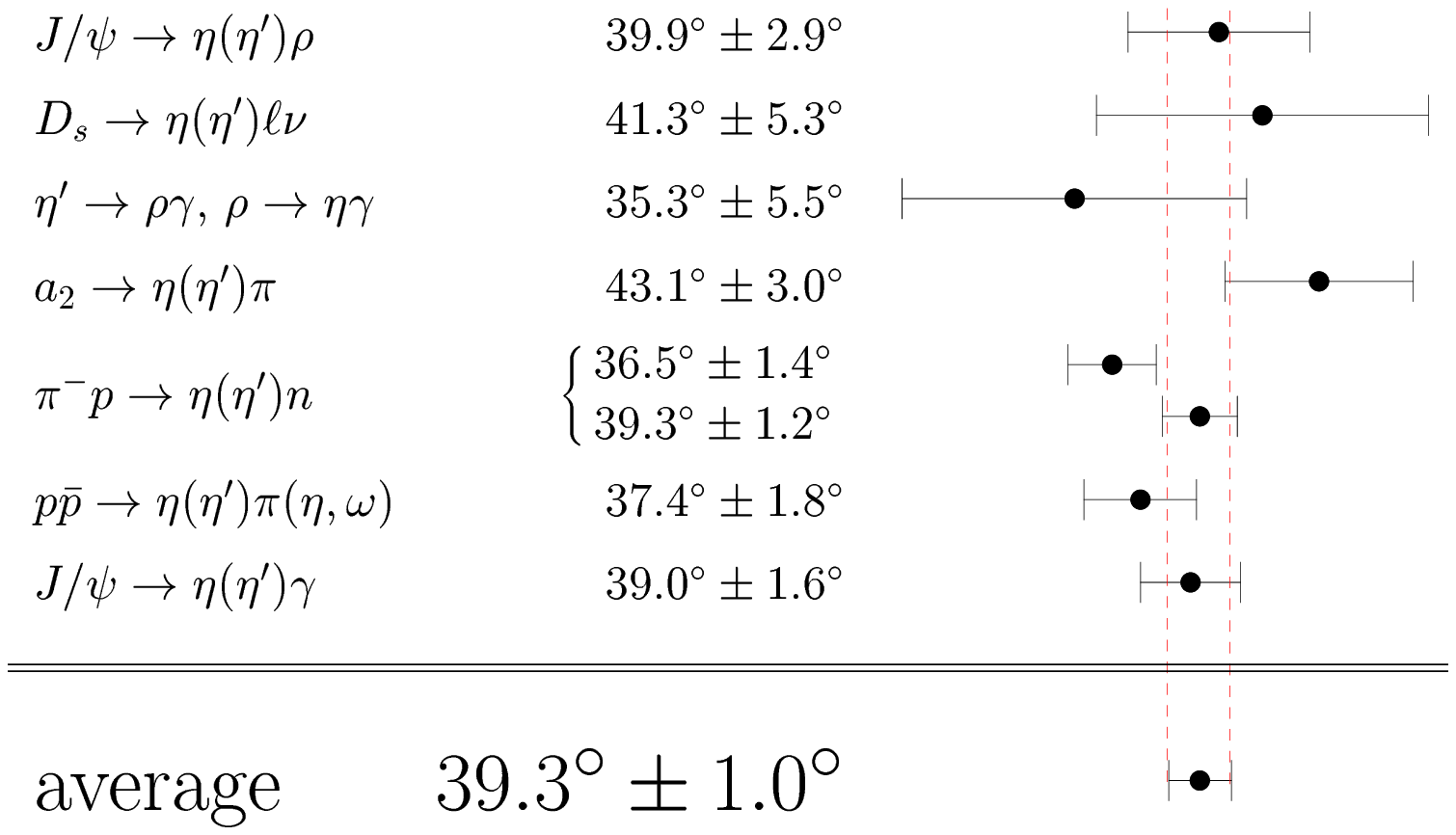 ,bb  = 70 495 520 750,  width=10cm}}}
\end{center}
\caption{Determination of the mixing angle $\phi$ from
different experimental processes, according to 
Ref.~\cite{FKS1,fel00} and references therein.}
\label{tab:phi}
\end{table}

Using the mass matrix (\ref{qsmass}) and the phenomenological value 
of $\phi$, one can evaluate phenomenological values for $a^2$ and
$y$. Finally, the two-photon decays of the $\eta$ and the $\eta'$
provide information on the decay constants. The PCAC results for the 
two-photon decay  widths of the $\eta$ and $\eta'$ in terms of $\phi$, 
$f_{q}$ and $f_{s}$ can be written as \cite{fel97},
\be
\Gamma[P\phantom{'}\to\gamma\gamma] =
\frac{\alpha^2}{32 \pi^3} \, \frac{M_{P}^3}{(\tilde{f}_\eta^{\rm eff})^2 }  \,.
\label{eq:gammapred}
\ee
Here the effective decay constants which are only
operative for the two-photon decays, are defined by
\be
\frac{1}{\tilde{f}_\eta^{\rm eff}} \equiv
\frac{1}{C_\pi} \left[C_q \, \frac{\cos\phi}{ f_{q}} -
      C_s \, \frac{\sin\phi}{f_{s}} \right] \ ; \qquad
\frac{1}{\tilde{f}_{\eta'}^{\rm eff}} \equiv
\frac{1}{C_\pi} \left[ C_q \, \frac{\sin\phi}{f_{q}} +
      C_s \, \frac{\cos\phi}{f_{s}}\right] \ , 
\label{eq:gammapredeff}
\ee
where $C_\pi = 1/(3 \sqrt2)$, $C_{q}=5/(9\sqrt2)$ and $C_{s}=1/9$ 
are electrical charge factors. The basic decay constants $f_q$ and 
$f_s$ can be evaluated from (\ref{eq:gammapred}) using the
experimental values for the decay width \cite{PDG}. The results are 
combined with the value of $y$ that we obtain from Eq.~(\ref{mix-par})
in order to improve the accuracy of the value for $f_s$. The
so-obtained results are also listed in Tab.~\ref{tab:par}. The 
phenomenological values for $\phi$, $f_q$ and $f_s$ provide
$ \tilde{f}_\eta^{\rm eff} = 1.02 \, f_\pi$ and 
$\tilde{f}_{\eta'}^{\rm eff} = 0.79 \, f_\pi$. As can be noticed from 
Tab.~\ref{tab:par} there is no substantial deviation between the 
theoretical and the phenomenological set of parameters, i.e.\ again 
higher-order flavor-symmetry breaking corrections etc., absorbed in 
the phenomenological values, seem to be reasonably small. A recent 
lattice calculation \cite{michael} yielded $y_{\rm lattice} \simeq
0.71$ which is in reasonable agreement with our values, too. 


\subsection{The mixing parameters in the octet-singlet basis}
 
Transforming the non-strange and strange axial-vector currents to
octet and singlet ones, one can connect the octet-singlet decay
constants defined in (\ref{dec}) to $f_{q}$ and $f_{s}$ with the
result \cite{FKS1}  
\begin{eqnarray}
&& f_8 = \sqrt{1/3 \, f_{q}^2 + 2/3 \, f_{s}^2} \ , \qquad
   \theta_8 = \phi - \arctan (\sqrt{2}\,f_{s}/f_{q})  \ , \cr
&& f_0 = \sqrt{2/3 \,f_{q}^2 + 1/3 \, f_{s}^2}  \ , \qquad
   \theta_0 = \phi - \arctan (\sqrt{2}\,f_{q}/f_{s}) \,, 
\label{fleu}
\end{eqnarray}
and thus 
\be
         \tan (\theta_0-\theta_8) = \sqrt{2}/3 \, 
(f_{s}/f_{q} - f_{q}/f_{s}) ~ .
\label{theta}
\label{anglediffex}
\ee
It is easy to convince oneself that for $\theta_8 \neq \theta_0$ the
basic states $\eta_8$ and $\eta_0$ defined through (\ref{osb}) are not
pure states in the sense of Eq.\ (\ref{opposite}); the matrix elements
$\langle 0 \mid J_{\mu 5}^{0(8)} \mid \eta_{8(0)}\rangle$ are non-zero
as a consequence of $\sudrei$ violation ($f_q\neq f_s$). In fact,
defining decay constants analogously to (\ref{dec}), one finds
\be 
f_{\eta_0}^8 = f_{\eta_8}^0 = \frac{\sqrt{2}}{3}\, (f_s - f_q)\,.
\label{wrong}
\ee

\begin{table}[thpb]
\begin{center}
{\small
\begin{tabular}{| ccccc || l |}
\hline
$\theta$ & $\theta_8$ & $\theta_0$ & $f_8/f_\pi$ & $f_0/f_\pi$ & method 
\\
\hline \hline
$-12.3^\circ$ & $-21.0^\circ$ & $-2.7^\circ$ & $1.28$ & $1.15$ & 
 $qs$--scheme (theo.\,) \cite{FKS1}
\\ 
$-15.4^\circ$ & $-21.2^\circ$ & $-9.2^\circ$ & $1.26$ & $1.17$ & 
 $qs$--scheme (phen.) \cite{FKS1}
\\
\hline
-- & $-21.4^\circ$ & $-7.0^\circ$ & $1.37$ & $1.21$ &
 energy-dependent scheme \cite{Escribano:1999nh}
\\
-- & $-20.4^\circ$ & $-0.1^\circ$ & $1.36$ & $1.32$ &
 VDM \& phenomenology \cite{Benayoun:2000au}
\\
-- & $-20.5^\circ$ & $-4^\circ$ & $1.28$ & $1.25$ & 
 $\chi$PT \& phenomenology \cite{kai98}
\\ 
$-20^\circ$ & x & x & $1.30$ & $1.04$ & 
 GMO \& $2\gamma$ decays \cite{Venugopal:1998fq}
\\ 
-- & $-22.2^\circ$ & $-9.1^\circ$ & $1.28$&  $1.20$ & transition form
factors \cite{fel97}\\
$-15.5^\circ$ & -- & -- & -- & -- & 
 phenomenology \cite{Bramon97} 
\\ 
$-12.6^\circ$ & $[-19.5^\circ]$ & $[-5.5^\circ]$ & $[1.27]$ & $[1.17]$ & 
 quark model  \cite{Schechter:1993iz}
\\
$-(23^\circ\!-\!17^\circ)$ & x & x & $1.2-1.3$ & $1.0-1.2$ & 
 phenomenology \cite{Dono,Gilman,BaFrTy95,GaLe85}
\\ 
$-9^\circ$ & $[-20^\circ]$ & $[-5^\circ]$ & $[1.2]$ & $[1.1]$ & 
anomaly \& meson masses \cite{Eides} \\
\hline
\end{tabular}}
\end{center}
\caption{Octet-singlet mixing parameters, 
evaluated in the quark-flavor scheme from the theoretical and 
phenomenological parameters given in 
Tab.~\ref{tab:par}, and comparison with other results. The values 
given in parentheses are not quoted in the original literature but 
have been evaluated by us from information given therein. Crosses 
indicate approaches where the difference between $\theta$, $\theta_0$ 
and $\theta_8$ has been ignored.}
\label{tab:comp}
\end{table}

In Tab.\ \ref{tab:comp} we show the results for the five mixing
parameters required in the octet-singlet basis, evaluated from $\phi$,
$f_q$ and $f_s$ and compare them to other results to be found in the
literature. Fair agreement can be observed between all approaches that
do not assume the equality of $\theta_8$ and $\theta_0$
\cite{Eides,kai98,FKS1,fel97,Escribano:1999nh,Benayoun:2000au,Schechter:1993iz}. 
Note that in some cases \cite{Escribano:1999nh,Benayoun:2000au}
different {\em parameterizations}\/ for the $\eta$--$\eta'$ mixing 
are given which lead, however, to the same physical results.
The analysis presented in \cite{Bramon97} basically determines the 
mixing angle $\phi$ along the same lines as in \cite{FKS1}. It 
therefore leads to a value for the mixing angle $\theta$ that is 
similar to our phenomenological one. In \cite{Bramon97} flavor 
symmetry breaking is encoded in the constituent quark masses instead 
of decay constants but the size of the effects is similar in both cases. 
In the analyses \cite{Dono,Gilman,BaFrTy95,GaLe85,Venugopal:1998fq}, in which 
the differences between $\theta_8$, $\theta_0$ and $\theta$ are 
ignored, values for $\theta$ around $-20^\circ$ have been obtained.
This is close to our phenomenological value of 
$\theta_8$ but quite different from the values for $\theta$ and $\theta_0$. 
The reason for this will become clear in Sec.~\ref{ss:ano}.

The general parameterization (\ref{oldmix}) combined with (\ref{fleu})
and the theoretical estimate (\ref{phen6}) 
implies the following ``sum rules''
\begin{eqnarray}
  f_\eta^8 \, f_\eta^8 +  f_{\eta'}^8 \, f_{\eta'}^8 &=& f_8^2  
          \ \simeq \ \frac13\, (4 f_K^2 - f_\pi^2) \ ,
\label{eq:f8rel}
\\[0.1em]
   f_\eta^8 \, f_\eta^0 +  f_{\eta'}^8 \, f_{\eta'}^0  
                   &=& f_8 f_0 \sin(\theta_8-\theta_0)
                 \ \simeq \ - \frac{2 \sqrt2}{3} \, (f_K^2-f_\pi^2)
\ .
\label{eq:t8t1rel}
\end{eqnarray}
Leutwyler and Kaiser \cite{kai98} derived these relations within the
framework of chiral perturbation theory from assumptions on flavor
symmetry breaking similar to (\ref{phen6}) which are required in order
to fix the parameters of the chiral effective
Lagrangian. Eqs.~(\ref{eq:f8rel}),(\ref{eq:t8t1rel}) are not a
consequence of the dynamical content of chiral perturbation theory
\cite{bass}. It is to be noted that Eq.~(\ref{eq:t8t1rel})
appears to be rather sensitive to higher order flavor symmetry
breaking effects etc.\/, as the comparison with our phenomenological
values shows. This sensitivity is related to that of the 
mixing angle $\theta_0$, see Tab.~\ref{tab:comp}.
The singlet decay constants $f_P^0$ and, consequently, $f_q$
and $f_s$, are renormalization-scale dependent \cite{kai98}
while the ratio $y=f_q/f_s$ as well as all mixing angles are 
scale-independent. The anomalous dimension controlling the 
scale-dependence of $f_P^0$ is of order $\als^2$ and therefore leads
to tiny effects in the basis decay constants which we discard. 
The sum rule for $f_P^0$ analogous to
(\ref{eq:f8rel}) is to be modified in order to take into account
the scale dependence \cite{kai98}. This additional OZI-rule violating
effect is also neglected.

\subsection{Anomaly matrix elements}
\label{ss:ano}
We have already pointed out the crucial role of the anomaly in the mass
matrix (\ref{qsmass}). Vacuum-particle matrix elements of the anomaly
operator $G\widetilde{G}$, which is occasionally termed the
topological charge density, are also of importance in other
processes. Therefore, we list various anomaly
matrix elements evaluated in the quark-flavor mixing scheme:
\ba
\langle 0 | \frac{\alpha_s}{4\pi} \, G \widetilde G | \eta_{8} \rangle
 & =& \sqrt{\frac23}\,\frac{f_q}{f_s\,} (f_s - f_q)\, a^2\,, \nn\\
\langle 0 | \frac{\alpha_s}{4\pi} \, G \widetilde G | \eta_{0} \rangle
 &=& \sqrt{\frac13}\, \frac{f_q}{f_s}\, (2 f_s + f_q)\, a^2\,, \nn\\
\langle 0 | \frac{\alpha_s}{4\pi} \, G \widetilde G | \eta\phantom{'}\rangle
 & =& - \sin \theta_8\, \frac{f_q}{f_s}\, \sqrt{2 f_s^2 + f_q^2 }\, a^2\,,\nn\\ 
\langle 0 | \frac{\alpha_s}{4\pi} \, G \widetilde G | \eta' \rangle
 &=& \phantom{-} \cos \theta_8\, \frac{f_q}{f_s}\, \sqrt{2 f_s^2 + f_q^2}\, a^2\,.
\label{eq:anomaly}
\ea
The corresponding matrix elements for $\eta_q$ and $\eta_s$ can be 
obtained from (\ref{aadef}) and (\ref{ydef}).

A particular noteworthy result is the non-vanishing of the
vacuum--$\eta_8$ matrix element; only in the limit of exact flavor
symmetry, $f_q=f_s$ it becomes zero. This again demonstrates the 
impurity of the $\eta_8$ state as defined in (\ref{osb}) with 
$\theta=\phi - \theta_{\rm ideal}$. Interesting is also the ratio of
the $\eta$ and $\eta'$ matrix elements
\be
\frac
{\langle 0 | \frac{\alpha_s}{4\pi} \, G \widetilde G | \eta\phantom{'}\rangle}
{\langle 0 | \frac{\alpha_s}{4\pi} \, G \widetilde G | \eta' \rangle}
       = -\tan \theta_8\,,
\label{ratio8}
\ee
The ratio of the anomaly matrix elements can also be expressed in
terms of the mixing angle $\phi$ and the masses of the physical
mesons. Using (\ref{qsb}), (\ref{aadef}) and (\ref{ydef}), one finds
\be
\frac
{\langle 0 | \frac{\alpha_s}{4\pi} \, G \widetilde G | \eta\phantom{'}\rangle}
{\langle 0 | \frac{\alpha_s}{4\pi} \, G \widetilde G | \eta' \rangle}
       = \left(\frac{M_\eta}{M_\eta'}\right)^2\, \cot \phi\,.
\ee
This relation has already been obtained by Ball et al.~\cite{BaFrTy95}
independent of the quark-flavor mixing scheme. However, its connection
with the mixing behaviour of the decay constants has not been
recognized in \cite{BaFrTy95}.

In the simple octet-singlet mixing scheme, defined by (\ref{osb}),
(\ref{oldmix}), one implicitly assumes that the anomaly only mediates
vacuum--$\eta_0$ transitions and that
${\langle 0 | G \widetilde G | \eta_8\rangle}=0$. These assumptions
imply the replacement of $\theta_8$ in (\ref{ratio8}) by $\theta$. 
Hence, analyses that are based on the assumption 
${\langle 0 | G \widetilde G | \eta_8\rangle}=0$, 
typically lead to large values of the mixing angle $\theta$ which
resemble that for $\theta_8$ (e.g., \cite{Venugopal:1998fq}) and are
in conflict with those values obtained from processes that only
involve state mixing. Examples of such analyses are that of the
Gell-Mann-Okubo formula or that of the decays
$\jp\to\gamma\eta,\,\gamma\eta'$. The latter process will be discussed in
Sect.~\ref{ss:rad} while a detailed investigation of the
Gell-Mann-Okubo formula in the light of the new ideas on
$\eta$--$\eta'$ mixing, can be found in \cite{FKS2,fel00}.  


\section{Masses and mixing angles versus $a^2$ and $y$}
\label{sec:a2}

\begin{figure}[t]
\begin{center}
 \psfig{file=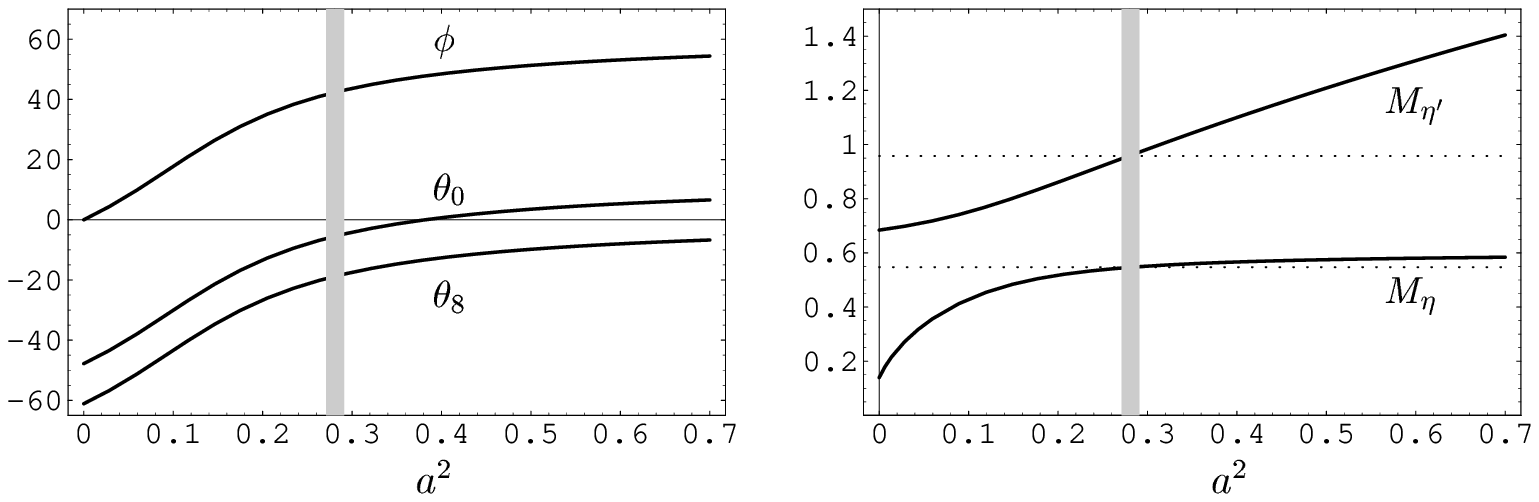, bb = 65 605 515 760, width=0.95\textwidth}
\end{center}

\caption{The mixing angles $\phi$, $\theta_8$ and
$\theta_0$ and the masses $M_\eta$ and $M_{\eta'}$ vs. the strength of
the anomaly parameter $a^2$ (for $y=0.78$ and quark mass terms
according to (\ref{masses})). The hatched vertical band
refers to the range $a^2 = 0.281 \pm 0.01$~GeV$^2$. The dotted lines
indicate the physical $\eta$ and $\eta'$ masses. $a^2$ in $\gev^2$ and
masses in GeV.}
\label{fig:a2}
\end{figure}

The investigations presented in
Sect.~\ref{sec:qsmix} clearly reveal the important role of $\sudrei$\/
symmetry breaking and the $\uaeins$ anomaly in understanding and
parameterizing $\eta$--$\eta'$ mixing. 
To elucidate further the role of the anomaly we consider the strength
of $a^2$ as a free parameter and evaluate the mixing
angles and the $\eta$, $\eta'$ masses for given values of $a^2$ by
diagonalizing the mass matrix (\ref{qsmass}). The values of the quark
mass terms (\ref{miidef}), (\ref{masses}) and of the flavor symmetry breaking
parameter $y\; (=0.78)$ are kept fixed. The results are plotted in
Fig.~\ref{fig:a2}.

In the limit $a^2 \to 0$ the non-diagonal elements of the mass matrix
(\ref{qsmass}) become negligible, and the pure flavor states $\eta_q$
and $\eta_s$ are the mass eigenstates with the masses
$M_\eta=m_{qq}\simeq M_\pi$ and $M_{\eta'}=m_{ss}\simeq \sqrt2 M_K$.   
The mixing angle $\phi$ tends to zero and, correspondingly, $\theta
\to - \theta_{\rm ideal}$. The substantial difference
between $\theta_8$ and $\theta_0$ remains constant, according to
(\ref{anglediffex}). For small values of $a^2$ the
situation is similar to the case of vector mesons where mixing is only
due to the weak, gluon-mediated, OZI-rule violating 
$q_i\bar{q}_i \to q_j\bar{q}_j$
transitions which are not enhanced by the $U(1)_A$ anomaly. 
One thus has almost ``ideal mixing'' between $\omega$ and $\phi$
mesons with a mixing angle $\phi_V$ of only about $3.4^\circ \pm
0.2^\circ$ as determined, for instance, from the ratio of the
$\phi\to\pi^0\gamma$ and $\omega\to \pi^0\gamma$ branching ratios
\cite{fel-kroll00}.   

Now, if $a^2$ is increased, $\eta_q$--$\eta_s$ mixing becomes stronger. 
At $a^2\simeq (0.26 - 0.28)~\gev^2$ the meson masses acquire their
physical values and $\phi$ is about $40^\circ$; this is the physical
region. If $a^2$ is amplified
further and becomes much larger than the quark
mass terms in (\ref{masses}), the mass matrix simplifies to
\begin{equation}
{\mathcal M}^2_{qs} \ \stackrel{a^2 \gg m_{ss}^2}{\;\longrightarrow\;} \ 
a^2\, \left(\matrix{
             2  & \sqrt{2}y \vspace{0.3em}\cr
                 \sqrt{2} y &  y^2 } \right)\,.
\label{limit}
\end{equation} 
Diagonalization of this mass matrix yields $M_{\eta'} \to \sqrt{2+y^2}\,
a$ while $M_\eta$ stays close to its physical value. For the mixing
angles one finds $\phi = \arctan[\sqrt2/y]$ and $\theta_8=0$ while
$\theta_0$ becomes positive. 
Although gluons are flavor-blind, mixing between different 
states is flavor-dependent because also the decay constants
$f_i$ are involved c.f.\ (\ref{ydef}).
On the other hand, if we now consider the additional limit
of exact $\sudrei$ symmetry, i.e.\ $y\to 1$, 
one obtains a ``democratic'' mass matrix where all elements in
(\ref{limit}) having the same strength  (apart from trivial factors 
of $\sqrt 2$ arising from the definition of the $\eta_q$ and
$\eta_s$ basis states).  Diagonalization of the
mass matrix (\ref{limit}) then provides a massles $\eta$ Goldstone meson,
and a heavy $\eta'$ meson with  $M_{\eta'} \to \sqrt3\, a$. 
For the mixing angles one finds $\phi\to \theta_{\rm ideal}$ and 
$\theta \simeq \theta_8 \simeq \theta_0 \to 0$, 
i.e.\ in this case the octet-singlet basis becomes the physical one.

\begin{figure}[thb]
\begin{center}
 \psfig{file=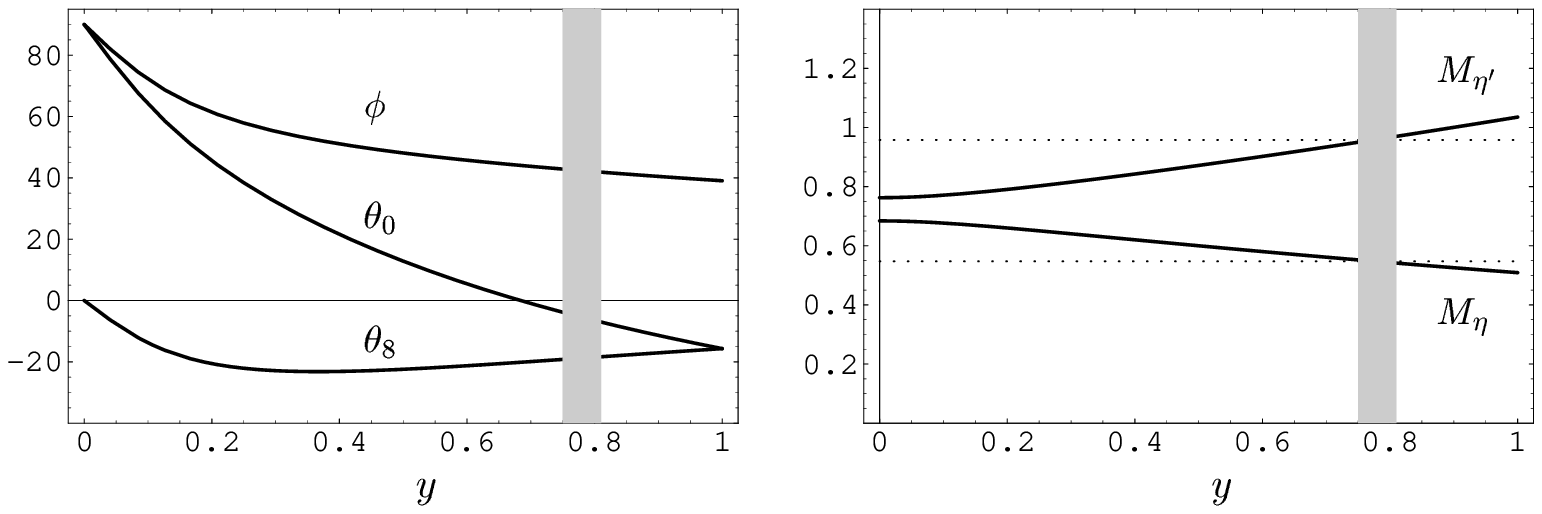, bb = 65 605 515 760, width=0.95\textwidth}
\end{center}

\caption{The mixing angles $\phi$, $\theta_8$ and
$\theta_0$ and the masses $M_\eta$ and $M_{\eta'}$ vs.\ $y$
(for $a^2=0.281$~GeV$^2$ and quark mass terms
according to (\ref{masses})). The hatched vertical band
refers to the range $y = 0.78 \pm 0.03$. The dotted lines
indicate the physical $\eta$ and $\eta'$ masses. All masses in
GeV.}
\label{fig:y}
\end{figure}

The results of a similar analyis of the mass matrix (\ref{qsmass})
where now the flavor symmetry breaking parameter $y$ is varied but the
anomaly parameter $a^2$ is kept fixed, are shown in Fig.~\ref{fig:y}.
In the limit $y\to 1$ one finds $\theta=\theta_8=\theta_0 = \phi -
\theta_{\rm ideal}$. Both the (simple) octet-singlet mixing scheme, (\ref{osb})
and (\ref{oldmix}), and the quark-flavor one, (\ref{qsb}) and
(\ref{phimix}), hold and are fully equivalent. If $y$ is getting
smaller than 1 the splitting between the three mixing angles
$\theta$, $\theta_8$ and $\theta_0$ sets in and becomes maximal
in the (academic) limit $y\to 0$. 

\section{Applications}
\label{sec:appl}
\subsection{Radiative decays of vector mesons}
\label{ss:rad}
\begin{figure}[t]
 \begin{center}
\psfig{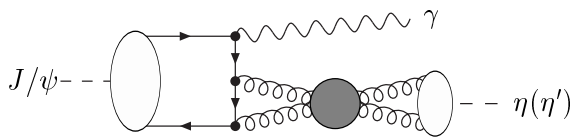}
   \end{center}
\caption{The decay $J/\psi \to \gamma\eta(\eta')$
        through the $\uaeins$ anomaly (indicated by the grey blob).}
\label{fig:jp}
\end{figure}

According to \cite{novikov} the radiative decays of heavy $S$-wave quarkonia
into $\eta$ or $\eta'$ proceed through the emission of the photon from the
heavy quark which subsequently annihilates into lighter quark pairs
through the effect of the anomaly, see Fig.\ \ref{fig:jp}. 
Making use of (\ref{eq:anomaly}), one finds for the ratio of decay
widths in the quark flavor scheme \cite{FKS1}
\be
R({}^3S_n) = \frac{\Gamma({}^3S_n \to\gamma\eta')} 
           { \Gamma({}^3S_n \to \gamma\eta)}
=\cot^2 \theta_8 \, \left(\frac{k_{\gamma\eta'}}{k_{\gamma\eta}}\right)^3
\label{rv}
\ee 
where $k_{12}=\sqrt{(M^2-m_1^2-m_2^2)^2-4m_1^2m_2^2}/(2M)$
denotes the final state's three-moment\-um in the rest frame of the
decaying particle. From the experimental value of $R(\jp)$ \cite{PDG}
one evaluates $|\theta_8|=22.0^\circ \pm 1.0^\circ$ or
$\phi=39.0^\circ\pm1.6^\circ$; the latter value has been used in the
phenomenological determination of the basic mixing parameters
\cite{FKS1}, see Tab.\ \ref{tab:phi}. From the phenomenological value
of $\theta_8$ quoted in Tab.\ \ref{tab:comp}, 
we predict $R(\psi')=5.8$ and $R(\Upsilon)=6.5$. The result for
$R(\psi')$ agrees with the experimental value $2.9 ^{+5.4}_{-1.8}$,
whithin the uncomfortably large errors \cite{Bai:1998ny}.
We repeat -- in the simple octet-singlet mixing scheme, defined by
(\ref{osb}) and (\ref{oldmix}), one would have interpreted $\theta_8$ in
(\ref{rv}) as $\theta$. 

One may also consider radiative transitions $\eta$ or $\eta'$ and
light vector mesons. In this case the $\uaeins$ anomaly does not
contribute but, as an additional complication, vector meson mixing is
to be taken into account. The recent measurement of the radiative
$\phi$ meson decays performed by KLOE \cite{KLOE} allows for another severe
test of the quark-flavor mixing scheme now. The ratio of the
corresponding decay widths, which is indeed measured by KLOE, is
given by
\be
R(\phi)= \cot^2{\phi}\,
      \left(\frac{k_{\eta'\gamma}}{k_{\eta\gamma}}\right)^3\,
         \left[ 1 -2 \, \frac{f_s}{f_q}\,
                      \frac{\tan{\phi_V}}{\sin{2\phi}}\right]\,,
\ee
from which we predict $R(\phi)= ( 5.66\pm 0.20 )\cdot 10^{-3}$ (for
$\phi_V=3.4^\circ\pm0.2^\circ$) in 
agreement with the KLOE result of $(5.30\pm 0.5\pm0.4) \cdot 10^{-3}$.
This analysis can be extended to the case of $\omega$ and $\rho$
mesons \cite{FKS2} but the quality of the experimental data for these
decays \cite{PDG} needs improvement before the mixing scheme can be 
examined seriously.


\subsection{The pseudoscalar meson photon transition form factor}

\label{sec:pgamma}

The transition form factors between pseudoscalar mesons and photons at
large momentum transfer are subject of intense theoretical interest,
see e.g.\ \cite{fel97,kro96,bro98,DKV}. Since the form factors are
sensitive to the decay constants they also provide a crucial test of
the quark-flavor mixing scheme. A leading twist analysis to
next-to-leading order (NLO) QCD accuracy, based on QCD factorization,
lead to (for $P=\pi^0, \eta, \eta'$)   
\ba
F_{P\gamma^*}(\qb,\omega) &=& 
\frac{\hat{f}_P^{\rm eff}}{3\sqrt{2}\, \qqb}\,
\int_{-1}^{\;1} {d} \xi\, 
\frac{\Phi_P(\xi,\mu_F)}{1-\xi^2\omega^2}\, 
\left[1 + \frac{\als(\mu_R)}{\pi}\,{\cal K}(\omega,\xi,\qb/\mu_F) 
\right] \nn\\[0.5em]
  &&+ \ \mbox{gluonic contribution for $ \eta$, $\eta'$} \ ,
\label{fpgvirtual}
\ea
where $\qqb = (Q^2 + Q'^2)/2$ is the average of the two photon
virtualities and $\omega= (Q^2 - Q'^2)/(Q^2 + Q'^2)$ 
the normalized difference. $\Phi_P(\xi,\mu_F)$ are the light-cone 
distribution amplitudes where $\xi$ is related to the parton momentum
fractions and $\mu_F$ is the factorization scale,
contain the relevant non-perturbative input.
For the case of the $\eta$ and $\eta'$ we
employ the quark-flavor mixing scheme. Eq.~(\ref{fpgvirtual}) is then
to be understood as a suitable superposition of the corresponding
expressions for the $\eta_q\gamma$ and $\eta_s\gamma$ transition form
factors. To simplify matters it is, in agreement with experiment
\cite{fel97}, assumed that the $\eta_q$ and $\eta_s$ distribution
amplitudes are approximately equal. The $\eta$--$\eta'$ mixing then
reflects itself in the effective decay constants, $\hat{f}_P^{{\rm
eff}}$, defined by:  
\be
   \hat{f}_{\eta\phantom{'}}^{{\rm eff}} \equiv 
   \frac{1}{C_\pi} \left[ C_q \, f_q \cos{\phi}
                    - C_s \, f_s \sin{\phi} \right] \,, 
\qquad
    \hat{f}_{\eta'}^{{\rm eff}} \equiv 
    \frac{1}{C_\pi} \left[ C_q \, f_q \sin{\phi}
                    + C_s \, f_s \cos{\phi} \right] \,.
\label{feff}
\ee
where $C_\pi$, $C_q$, and $C_s$ are defined after (\ref{eq:gammapredeff}).
In case of the pion one simply has $\hat{f}_\pi^{\rm eff}=f_\pi$. 
The effective decay constants, $\hat{f}_P^{{\rm eff}}$, differ from
the $\tilde f_P^{\rm eff}$ defined in 
(\ref{eq:gammapredeff}) for the decays into two real photons. Their
numerical values ($\hat{f}_\eta^{\rm eff} = 0.98 \, f_\pi$,
and $\hat{f}_{\eta'}^{\rm eff} = 1.62 \, f_\pi$) are to be compared with
the values for $\tilde f_P^{\rm eff}$ quoted after (\ref{eq:gammapredeff}). 
In particular, $\hat{f}_{\eta'}^{\rm eff}$ is markedly different from 
$\tilde f_{\eta'}^{\rm eff}$ while {\em by accident}\/ one
has  $\hat{f}_{\eta}^{\rm eff} \simeq \tilde{f}_{\eta}^{\rm eff}$. 
This has lead to some confusion in early
attempts to interpret the experimental data \cite{cleo98,acc98} for
the $P\gamma$ transition form factors in the real photon limit ($Q'=0$) 
in terms of simple interpolation formulas between
(\ref{eq:gammapredeff}) and (\ref{fpgvirtual}). This issue has been resolved in 
\cite{interpol} where also the proper interpolation formulas are
given.\footnote{Note that the question of how to describe the $Q^2$
dependence of the $P\gamma$ transition form factors also plays a role
in the determination of the hadronic light-by-light scattering
contribution to the muon anomalous magnetic moment which has received
a lot of attention in the last months \cite{gminus2}.}

\begin{figure}[th]
\parbox{\textwidth}{
\begin{center}
\includegraphics[width=4.4cm, bb=200 470 440 660,clip=true]{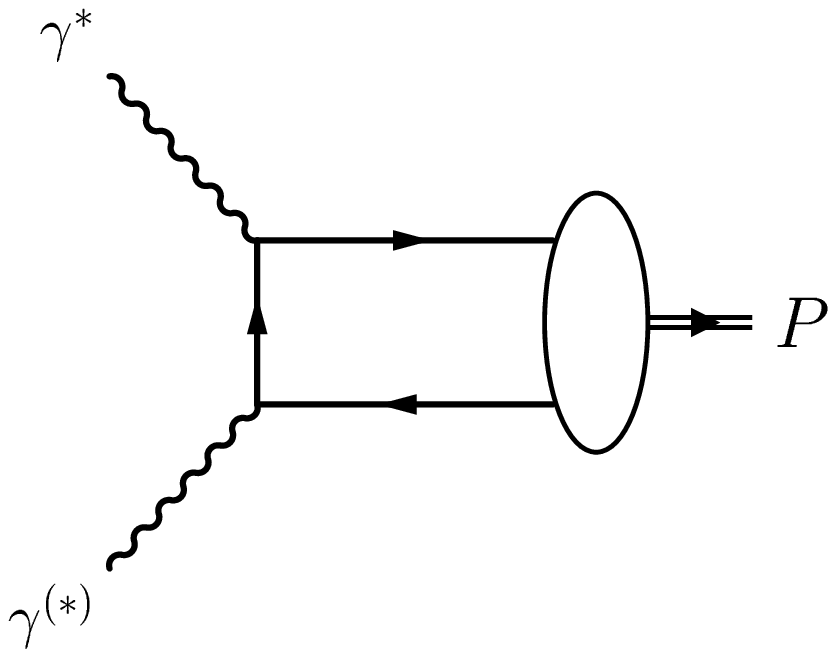}
\hspace{0.5em}
\includegraphics[width=4.6cm, bb=200 460 482
660,clip=true]{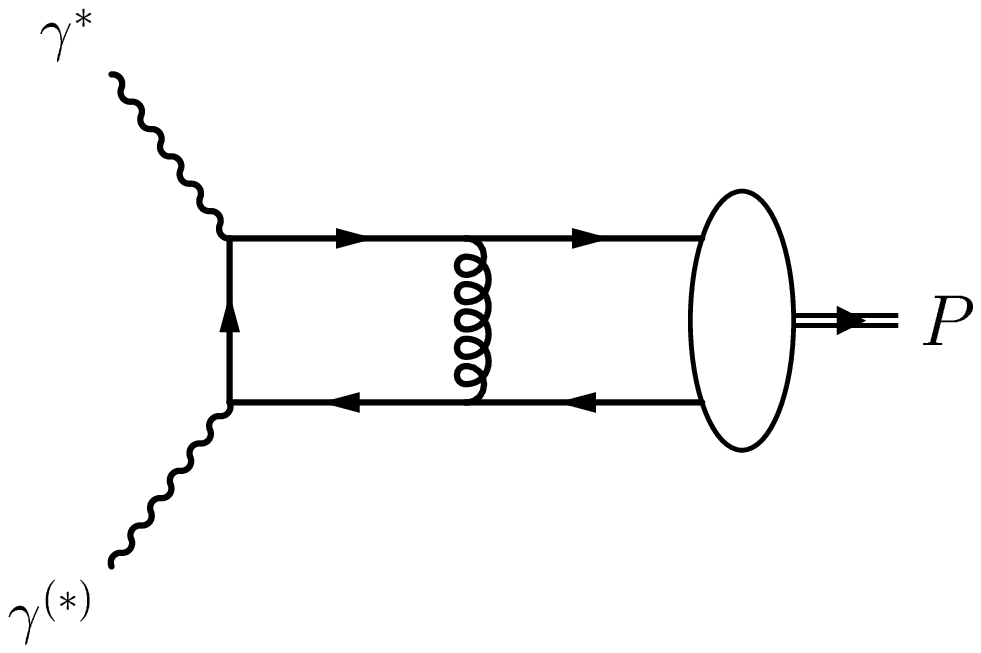}
\hspace{0.5em}
\includegraphics[width=4.4cm, bb=200 470 480 660] {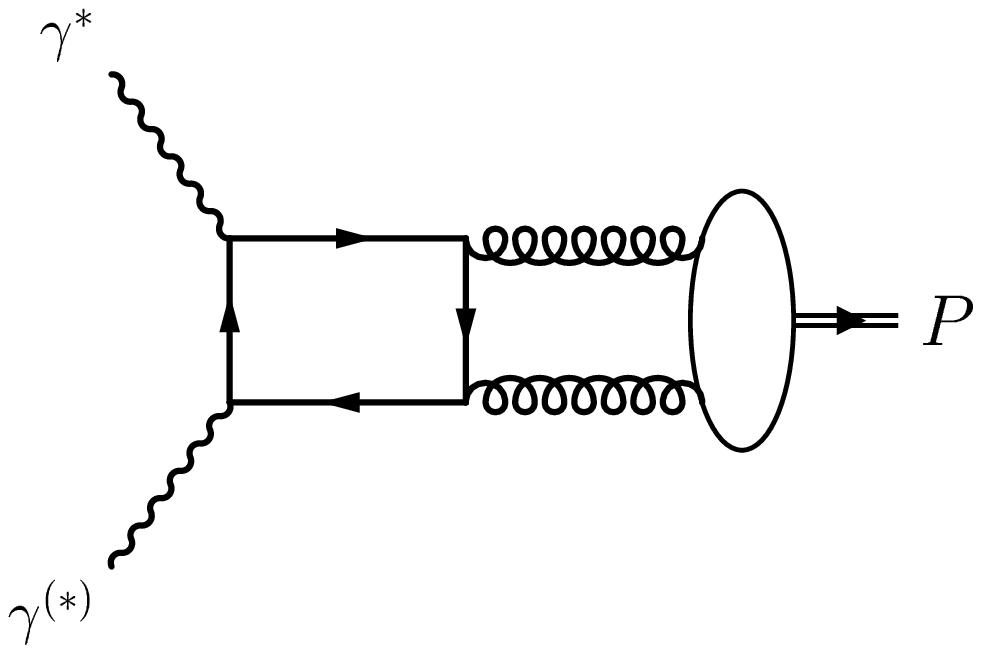}
\end{center}}
\caption{Sample Feynman graphs contributiong to meson-photon
transition form factor.}
\label{fig:graph}
\end{figure}
Sample Feynman graphs contributing to the form factor at NLO are shown
in Fig.~\ref{fig:graph}. Here, ${\cal K}$ is a known function
evaluated from the NLO graphs. To NLO there is also a contribution
from the leading-twist two-gluon distribution amplitudes of $\eta$ and
$\eta'$. These two-gluon distribution amplitudes mix with the quark
singlet distributions under evolution \cite{singlet}. Their size is
not yet clear \cite{fel97,ali}, but in the case of the transition form
factors, where the gluonic contributions only enters at NLO, they seem to
be negligible. This is consistent with the neglect of OZI-rule
violating contributions. It is important to realize that the $\uaeins$ anomaly
also contributes to the two-gluon Fock states but it generates
higher-twist distribution amplitudes \cite{novikov} which are
suppressed by inverse powers of the large scale, $\qqb$. Thus, despite
the fact that anomaly matrix elements $\langle 0 | G \tilde G |
P(p) \rangle \propto a^2$ are large, the gluonic content of the $\eta$
and $\eta'$ mesons are hardly perceptible in the transition form
factors at large momentum transfer. For low scales, on the other hand,
the suppression of higher-twist contributions is not operative and the
$\uaeins$ anomaly plays an important role as we discussed in this article. 
In cases where the leading-twist contributions are strongly suppressed
for one or the other reason the higher-twist gluonic contribution may
be dominant even at large scales. An example is set by the OZI-rule
forbidden $\jp\to\gamma\eta,\,\gamma\eta'$ decays which, although
taking place at the large scale $\mjp^2$, are anomaly-controlled 
(see Sect.~\ref{ss:rad}).   

One can show \cite{DKV} that for $\omega \lsim 0.6$ and $\qb \gsim  2
\gev$ the form factors become independent of the form of the distribution
amplitudes to a high degree of accuracy which leads to
\be
F_{P\gamma^*\gamma^*} \ \stackrel {\omega \leq 0.6}
{\;\longrightarrow\;} \ \frac{\sqrt{2} \hat f_P^{\rm eff}}{3 \overline{Q}^2}
[1-\frac{\alpha_s}{\pi}] + {\cal O}(\omega^2)\,.
\label{limita}
\ee
In this kinematical region we thus have a parameter-free prediction from QCD to
leading-twist accuracy. It has a status comparable to the famous
expression of the cross section ratio $R=\sigma(e^+e^-\to
{\rm hadrons})/\sigma(e^+e^-\to \mu^+\mu^-)$ or to that of the Bjorken sum
rule. Hence, (\ref{limita}) well deserves experimental verification.

\begin{figure}[tbhp]
\parbox{\textwidth}{
\begin{center}
\includegraphics[width=7.8cm, bb=20 30 590 645,angle=-90,clip=true]{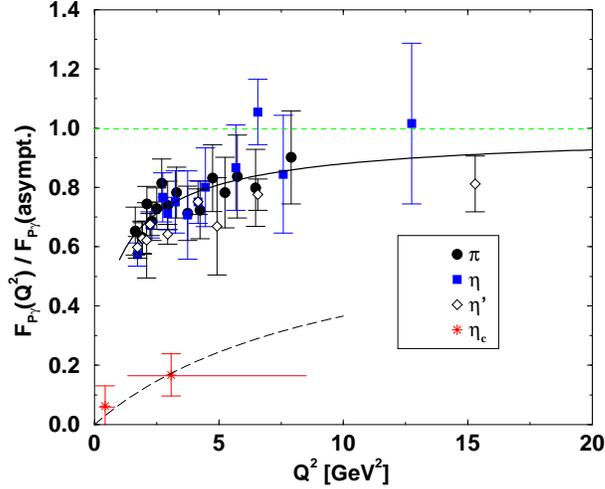}
\end{center}}
\caption{The $P\gamma$ transition from factors scaled by the LO
asymptotic result. Data are taken from \cite{cleo98,acc98,L3-etac},
the theoretical results from \cite{fel97} (solid lines) and
\cite{fel97a} (dashed-dotted line).}
\label{fig:ff}
\end{figure}

In the limit of one real photon ($Q'=0$) while the other one is highly
virtual photon ($Q^2\to\infty$) the form factors
become independent of the shape of the distribution amplitude, too,
since any distribution amplitude evolves into the asymptotic form
$\phi_P\to \phi_{AS}=3/2 (1-\xi^2)$ and $\phi_P^{g}\to 0$.
In this limit one finds
\be F_{P\gamma} \stackrel{Q^2\to\infty}{\;\longrightarrow\;} 
               \frac{\sqrt{2} \hat f_P^{\rm eff}}{Q^2}\, 
                    \left[1-\frac53\frac{\als}{\pi}\right]\,.
\ee
In Fig.\ \ref{fig:ff} the data for the light pseudoscalar
meson-photon transition form factors \cite{cleo98,acc98}, 
scaled by the leading-order (LO) asymptotic results,  are
shown. We see that the data for $\pi$, $\eta$ and $\eta'$ form factors
are universal within the errors and lie about $20\%$ below the LO
asymptotic value. There are many attempts to explain this difference
\cite{fel97,kro96,bro98,DKV}. The $\als$
corrections, choosing $\qb=Q/\sqrt{2}$ as the renormalization scale,
account for about a half of the deviation. Utilizing a distribution
amplitude that is somewhat narrower than the asymptotic one (in terms
of the Gegenbauer coefficients: $B_2=-0.06$, $B_n=0$ for $n\geq 4$ \cite{DKV}),  
the NLO leading-twist result is in agreement with experiment. Lower
renormalization scales require distribution amplitudes even closer to
the asymptotic form \cite{bro98}. Other possibilities to explain the
difference between experiment and the asymptotic result are the
inclusion of transverse degrees of freedom, e.g.\ \cite{fel97} -- these
results are shown in Fig.~\ref{fig:ff} -- or higher-twist corrections. 
The theoretical results for the three form factors agree well with
experiment.
The analysis of the transition form factors and the successful
comparison with experiment tests not only our understanding of the
meson distribution amplitudes but also the mixing behaviour of the
$\eta$ and $\eta'$ decay constants \cite{fel97}.

For comparison we also show in Fig.\ \ref{fig:ff} the $\eta_c \gamma$
transition form factor. Obviously, this form factor behaves differently. 
The reason is clear -- there is a second large scale in this process, 
namely the $\eta_c$ mass which leads to the observed suppression 
\cite{fel97a}. If the measured values of $Q^2$ were 
so large that the $\eta_c$ mass  could be neglected as compared
to it, this form factor would exhibit universality, too.

\subsection{The $\eta$, $\eta'$--nucleon coupling constants}

The coupling constants of the pseudoscalar mesons with nucleons are
important ingredients in many analyses of hadronic reactions at low
energies. In order to obtain an estimate of these coupling constants,
we use generalized Goldberger-Treiman (GT) relations,
and apply again the quark-flavor scheme
\cite{fel00}
\ba
  2 M_N G_A^3 &=& f_\pi \, g_{\pi NN}\,, \nn\\[0.3em]
  2 M_N G_A^a &=& \sum_{P=\eta,\eta'} f_P^a \, g_{PNN}\,, \qquad a =0,8\,.
\label{GT}
\ea
The $G^a_A$ are the axial vector coupling constants and
$M_N$ is the nucleon mass. The GT relations differ from the
Shore-Veneziano ansatz \cite{shore} where an additional direct
coupling of the Veneziano ghost to the nucleon has been allowed for.
The axial-vector couplings are known from phenomenology:
\ba
G_A^3 &=& 0.900\pm 0.002 \qquad {\rm from\; neutron\; \beta-decay}\,,\nn\\
G_A^8&=& \phantom{0} 0.24 \pm 0.01\phantom{0} \qquad {\rm from\; hyperon\;
                \beta-decay}\,,\nn\\
G_A^0 &=& \phantom{0} 0.16 \pm 0.10 \phantom{0} \qquad {\rm from\;
Bjorken\; sum\; rule}\,. 
\ea
The latter result holds at a scale of 5 GeV$^2$ \cite{SMC}.
For the case of the pion the phenomenological results for the
axial-vector couplings lead to $g^2_{\pi NN}/(4\pi)=13.2$, a 
value that is somewhat smaller than that one obtained from dispersion
theory \cite{ericsson} ($=14.2$). Potential models \cite{swart} and
nucleon-nucleon scattering phase shift analysis \cite{pawan} provide
values which are closer to the GT value. The origin of this little
discrepancy between the GT result and phenomenology is not yet
clear. 
The GT relation (\ref{GT}) also provides
\be
g^2_{\eta NN}/(4\pi) = 0.92\pm 0.3\,, \qquad  g^2_{\eta' NN}/(4\pi)=0.2\pm 0.2
\label{coupling}
\ee
A recent measurement of near-threshold $\eta'$ production in
proton-proton collisions \cite{moskal} provided the bound 
$g^2_{\eta' NN}/(4\pi)<0.5$. This admittedly model-dependent result  
is in agreement with the GT value (\ref{coupling}). A dispersion
analysis of the six nucleon-nucleon forward scattering amplitudes
\cite{grein} yielded $g^2_{\eta(\eta') NN}/(4\pi)<1.0$ in agreement
with the GT relation, too. The $\eta$ and $\eta'$ couplings could 
not be disentangled from each other in this analysis. One-boson
exchange potentials \cite{swart}, on the other hand, provide much
larger values ($g^2_{\eta NN}/(4\pi) = 3.7$, $g^2_{\eta'
NN}/(4\pi)=4.2$) which are in conflict with
(\ref{coupling}). Attention must be paid to the fact that in the OBE
potential models only meson exchange in a non-relativistic reduction 
is taken into account while contributions from multi-meson exchanges
are ignored. In the dispersion analysis \cite{grein}, on the other hand, 
evidence for contributions from the $2\pi$ and $3\pi$ continuum has
been found. With regard to this one should take coupling constants
obtained in potential  models with some care. They are rather
effective parameters than fundamental quantities.

\subsection{Isospin-singlet admixtures to the pion}

Isospin violation in the pseudoscalar meson sector can be viewed as
$\eta$ and $\eta'$ admixtures to the pion:
\be 
|\pi^ 0\rangle = \Phi_3 + \varepsilon |\eta \rangle + \varepsilon'
        |\eta'\rangle
\ee
where $\Phi_3$ denotes the pure isospin-triplet state. A straightforward
generalization of the quark-flavor mixing scheme by treating the $u$-
and the $d$-quark separately, allows for a determination of the
parameters $\varepsilon$ and $\varepsilon'$ \cite{FKS2} 
\be 
\epsilon  = \cos\phi \, \frac{m_{dd}^2 - m_{uu}^2}
                             {2 \, (M_\eta^2 - M_\pi^2)} 
\ , \qquad 
\epsilon'  = \sin\phi \, \frac{m_{dd}^2 - m_{uu}^2}
                             {2 \, (M_{\eta'}^2 - M_\pi^2)}\,.
\label{epsepsp} 
\ee
The quark mass difference $m^2_{dd} - m^2_{uu}$ can be estimated from 
$2 (M_{K^0}^2 - M_{K^\pm}^2 + M_{\pi^\pm}^2 - M_{\pi^0}^2)$ in which,
according to Dashen \cite{dashen}, masses of electromagnetic origin
are expected to cancel to a large extent. A possible difference in the
$u$ and $d$ quark decay constants is ignored in the derivation of 
(\ref{epsepsp}). With the phenomenological value of $39.3^\circ$ for
the mixing angle one obtains
\be
 \varepsilon = 0.014\,, \qquad  \varepsilon' = 0.0037\,.
\ee
An extraction of $\varepsilon$ from the anomaly dominated decays
$\Psi'\to \jp\,\pi^0,\; \jp\,\eta$
\be
          \frac{\Gamma(\Psi'\to\jp\,\pi^0)}{\Gamma(\Psi'\to\jp\,\eta)} = 
           \left(\frac{\varepsilon}{\cos^2\phi}\right)^2 \, 
                  \left(\frac{k_{\jp\,\pi^0}}{k_{\jp\,\eta}}\right)^3\,,
\ee
on the other hand, yields $\varepsilon=0.026\pm 0.003$. The observed
violation of charge symmetry in the cross sections for $\pi^+d\to
pp\eta$ and  $\pi^+d\to nn\eta$ \cite{tippens} lead to the same value
($\varepsilon=0.026\pm 0.007$) within a slightly model-dependent
analysis. Thus, it seems that there is a discrepancy of about a
factor of 2 whose origin is not yet clear. Is the $u-d$ mass
difference underestimated or are higher order electromagnetic or other
corrections lacking? The neglected OZI-rule violating effects are of
the same order as $\epsilon$, namely they amount to a few percent.
One may therefore suspect that the quark-flavor mixing
scheme, defined by (\ref{qsb}) and (\ref{phimix}), has reached its
limits of accuracy for effects of that size. It is possible that at
that level of precision three mixing angles, $\phi$, $\phi_q$ and
$\phi_s$, are required for an adequate description of all aspects of
$\eta$--$\eta'$ mixing.

Isospin violation play an important role in the analysis of the decays $\eta
(\eta') \to 3\pi$ \cite{leutwyler,ametller} as well as in the
investigation of CP-violation in $B\to\pi\pi$ decays \cite{gardner}.
They break the isospin triangle relation 
\be 
{\mathcal M}(\bar{B}^0\to\pi^0\pi^0) ={\mathcal M}(\bar{B}^0\to\pi^+\pi^-)/\sqrt{2} 
                  + {\mathcal M}(\bar{B}^-\to\pi^0\pi^-)\,,
\ee
and may therefore affect the determination of the CKM-angle $\alpha$.

One may wonder whether the $\eta$ and $\eta'$ admixtures to the $\pi^0$
do not affect the interpretation of the $\jp\to\gamma\pi^0$  decay.
This is, however, not the case as a quick estimate reveals. The
anomaly contribution to this process through these admixtures is 
\be
          \frac{\Gamma(\jp\to\gamma\pi^0)}{\Gamma(\jp\to\gamma\eta)} = 
            |\varepsilon - \cot \theta_8\, \varepsilon'|^2 \, 
                  \left(\frac{k_{\gamma\pi^0}}{k_{\gamma\eta}}\right)^3\,,
\ee
which provides to a value of only $6.1\cdot 10^{-4}$ for the ratio of
decay widths while the experimental value is $(4.5\pm 1.6)\cdot
10^{-2}$ \cite{PDG}. Thus, we can conclude - the radiative decay of the
$\jp$ into the $\pi^0$ is mediated by $\cbc\to \gamma \to \qbq$, a
contribution that is proportional to the $\pi\gamma$ transition form
factor, see Sect.~\ref{sec:pgamma}, and by a vector meson dominance
contribution $\jp\to\rho\pi^0\to\gamma\pi^0$ and not by the $\uaeins$
anomaly as the corresponding decays into $\eta$ or $\eta'$.

\section{Conclusion}
\label{sec:summ}

We have discussed $\eta$--$\eta'$ mixing in the octet-singlet and in
the quark-flavor basis. We have shown that for a complete and
consistent understanding it is not sufficient
to consider only state mixing. We therefore carefully considered the
mixing behaviour of the decay constants and of the matrix elements of
the $\uaeins$ anomaly operator. In this context,
the quark-flavor mixing scheme, defined through Eqs.~(\ref{qsb}) 
and (\ref{phimix}), appears to be favored since it only
requires three basic mixing parameters ($\phi$, $f_q$, $f_s$) to
describe all aspects of $\eta$--$\eta'$ mixing to an accuracy of about
a few percent, i.e.\ at a level at which OZI-rule violations become 
noticeable. The three basic mixing parameters can be determined 
with the help of the divergences of the axial-vector currents, which 
embody the $\uaeins$ anomaly, and first order $\sudrei$ symmetry 
relations. Alternatively, the mixing parameters can be determined
by using various phenomenological input. The most important
processes to verify the consistency of quark-flavor mixing scheme are 
the radiative decays of $S$-wave quarkonia into $\eta$ or $\eta'$,
and the transition form factors between photons and pseudoscalar
mesons.

On the other hand, the simple octet-singlet mixing scheme, defined by
Eqs.~(\ref{osb}) and (\ref{oldmix}), is obsolete and in clear conflict 
with phenomenology. Of course, one may still use the octet-singlet basis
but, if decay constants or matrix elements of the anomaly operator
are considered, one has to allow for two additional mixing angles 
$\theta_8$ and $\theta_0$, (see (\ref{newmix}),(\ref{ratio8})) with 
substantially different values.

The reason for the preference of the quark-flavor scheme is the
smallness of OZI-rule violations, which amount to only a few percent
as can be seen, for instance, from the difference between the angles 
$\phi_q$ and $\phi_s$ as determined in the analysis of the
$\eta\gamma$ and $\eta'\gamma$ transition form factors \cite{fel97}.
On the other hand, $\sudrei$ symmetry is broken at the level of 
$10 - 20 \%$, if one takes the difference between the pseudoscalar
decay constants as a relevant measure, and cannot be neglected, in
contrast to OZI-rule violations. Via Eq.~(\ref{anglediffex}) this
immediately rules out the simple ansatz (\ref{oldmix}).

We finally remark that a simple one-angle description of
$\eta$--$\eta'$ mixing can, at most, hold in one basis except
in the trivial case of perfect $\sudrei$ symmetry. 
In any other basis, 
mixing is unavoidably more complicated as can easily be shown 
by calculations analogous to the one sketched in Sect.~\ref{sec:qsmix}.

\section*{Acknowledgements}
P.K.\ thanks Goran F\"aldt for the kind invitation to the workshop on
Eta Physics and for the hospitality and the excellent organisation in Uppsala.


\end{document}